\documentclass[doublecol]{epl2} 

\title{A mixed SOC-turbulence model for nonlocal transport and space-fractional Fokker-Planck equation}
\shorttitle{A mixed SOC-turbulence model and space-fractional Fokker-Planck equation} 

\author{A.~V.~Milovanov$^{1,2}$ and J.~Juul~Rasmussen$^3$}

\shortauthor{A.~V.~Milovanov and J.~J.~Rasmussen}

\institute{$^1$Associazione EURATOM$-$ENEA sulla Fusione, Centro~Ricerche~Frascati, I-00044 Frascati, Rome, Italy

$^2$Department of Space Plasma Physics, Space Research Institute, Russian Academy of Sciences, 117997 Moscow, Russia

$^3$Association EURATOM$-$DTU, Physics Department, Technical University of Denmark, DK-4000 Roskilde, Denmark}


\abstract{The phenomena of nonlocal transport in magnetically confined plasma are theoretically analyzed. A hybrid model is proposed, which brings together the notion of inverse energy cascade, typical of drift-wave- and two-dimensional fluid turbulence, and the ideas of avalanching behavior, associable with self-organized critical (SOC) behavior. Using statistical arguments, it is shown that an amplification mechanism is needed to introduce nonlocality into dynamics. We obtain a consistent derivation of nonlocal Fokker-Planck equation with space-fractional derivatives from a stochastic Markovian process with the transition probabilities defined in reciprocal space.}

\begin{document}

\maketitle

\section{Introduction} In magnetically confined plasmas, perturbative experiments \cite{Mantica} with plasma edge cooling and heating power modulation reveal anomalously fast transport of edge cold pulses to plasma core, not compatible with major diffusive time scales \cite{Pulse,Pulse+}. It was argued that local transport models were problematic to accommodate the observed behaviors, and that there was a connection between the cold pulse problem and the phenomena of nonlocal transport described by kinetic equations with fractional derivatives in space \cite{Nature,Klafter,Report}. The latter are integro-differential operators generalizing the notion of ordinary differentiation to fractional orders \cite{Klafter,Samko}. Fluctuation-induced transport, as confirmed in a variety of different experiments, involves long-range dependence and non-Gaussian scaling, and the existing phenomenology has been discussed in terms of {\it ad hoc} equations with fractional modification of the Laplacian \cite{Pulse,Castillo2}. However, a detailed understanding of the underlying mechanisms driving nonlocal transport is still not at hand. In particular, it is not clear which key physics ingredients behind the plasma confinement one really needs in order to generate nonlocal behavior. Another important issue here is validation of nonlocal models from microscopic dynamics of diffusing charged particles. The goal of the present study is to obtain progress over these topics.  

In this work, we address the nonlocal transport problem from a more fundamental perspective, namely, by advancing the concept of nonlocal transport driven by a stochastic noise process of the L\'evy type. The key idea behind this approach is inspired by the early work of Chechkin and Gonchar \cite{Gonchar}, but with a different derivation using the notion of transition probability in reciprocal space. More so, we show that, when account is taken for boundary dissipation processes, and for associate nonlinear feedbacks in the plasma edge, the model leads to important implications with respect to the phenomena of self-organized criticality (SOC). The theoretical concept of SOC has been initially applied by Bak {\it et al.} \cite{Bak} to describe sandpile avalanches at a critical angle of repose, and has been generalized to nonlinear dissipative systems that are driven in a critical state. Self-organized criticality occurs through a nonlinear feedback mechanism \cite{Sor} triggering intermittent, avalance-like transitions between different metastable states. It has been argued theoretically, see, e.g., Refs. \cite{Newman,Carr1}, that transport processes in magnetically confined plasmas have some of the characteristics of SOC systems, and that the behavior is universal among the different confinement devices (tokamaks and stellarators) \cite{Carr1,Pedrosa99}. These topics are summarized in a recent book \cite{Ash2013}.

In what follows, we shall first derive a L\'evy-fractional Fokker-Planck equation from a generic Markovian stochastic process in configuration space. Then we shall attempt to set the model in a more general context and discuss the connections with turbulence- and SOC-associated phenomena. A milestone here is the idea of drift-wave turbulence (and, more generally, of two-dimensional fluid turbulence) subordinated to SOC. We suggest that the nonlocal character of the fractional derivative in space reflects the presence of large-amplitude, intermittent bursts in the flow; whereas the fractional order of the differintegration, a numerical fractional exponent used to generalize the derivatives, is decided by the overall SOC organization of the turbulent system near the state of marginal stability. The implication is that SOC and turbulence phenomena couple in magnetic confinement devices, and that they operate in concert to produce the expected \cite{Pulse,Castillo2} fractional form of the transport equations. Neither of these processes alone will be sufficient to introduce nonlocality to the transport. More so, our results suggest that the large bursts occur via a turbulent amplification process, which involves both the SOC-associated avalanches and the inverse energy cascade characteristic of drift-wave- and two-dimensional fluid turbulence. This amplification process will manifest itself in the form of algebraic tails on top of a typically log-normal-like probability distribution function of the flux-surface averaged transport. We consider this as the main theoretical prediction of the proposed model.    

\section{Fractional Fokker-Planck equation} We work with a Markovian (memoryless) stochastic process defined by the evolution equation\footnote{The case of velocity-space transport, though conceptually similar, is not discussed here. Nor do we discuss processes with trapping, leading to slow diffusion and fractional time derivatives in the end.}
\begin{equation}
f(x, t+\Delta t) = \int_{-\infty}^{+\infty} f(x-\Delta x, t) \psi (x, \Delta x, \Delta t)d\Delta x,
\label{1} 
\end{equation}
where $f (x, t)$ is the probability density of finding a particle (random walker) at time $t$ at point $x$ and $\psi (x, \Delta x, \Delta t)$ is the transition probability density of the process. Note that the ``density" $\psi (x, \Delta x, \Delta t)$ is defined with respect to the increment space characterized by the variable $\Delta x$. It may include a parametric dependence on $x$ as soon as inhomogeneous systems are considered. Here, for the sake of simplicity, we restrict ourselves to the homogeneous case, and we omit the $x$ dependence in $\psi (x, \Delta x, \Delta t)$ to obtain
\begin{equation}
f(x, t+\Delta t) = \int_{-\infty}^{+\infty} f(x-\Delta x, t) \psi (\Delta x, \Delta t)d\Delta x.
\label{2} 
\end{equation} 
Then $\psi (\Delta x, \Delta t)$ defines the probability density of changing the spatial coordinate $x$ by a value $\Delta x$ within a time interval $\Delta t$, independently of the running $x$ value. The integral on the right of Eq.~(\ref{2}) is of the convolution type. In the Fourier space this becomes
\begin{equation}
\hat f(k, t+\Delta t) = \hat f(k, t) \hat \psi (k, \Delta t),
\label{3} 
\end{equation} 
where the integral representation  
\begin{equation}
\hat \psi (k, \Delta t) = \hat \mathrm{F} \{\psi (\Delta x, \Delta t)\} \equiv \int_{-\infty}^{+\infty} \psi (\Delta x, \Delta t) e^{ik\Delta x} d\Delta x
\label{Fourier} 
\end{equation} 
has been used for $\hat \psi (k, \Delta t)$, and similarly for $\hat f(k, t)$. Letting here $k\rightarrow 0$, it is found that 
\begin{equation}
\lim_{k\rightarrow 0}\hat \psi (k, \Delta t) = \int_{-\infty}^{+\infty} \psi (\Delta x, \Delta t) d\Delta x.
\label{F2} 
\end{equation} 
The improper integral on the right hand side is nothing else than the probability for the space variable $x$ to acquire {\it any} increment $\Delta x$ during time $\Delta t$. For memoryless stochastic processes without trapping, this probability is immediately seen to be equal to 1 (i.e., the diffusing particle takes a displacement anyway in any direction on the $x$-axis), given that the time interval $\Delta t$ is longer than the characteristic width of the driving-force spikes. Thus,
\begin{equation}
\lim_{k\rightarrow 0}\hat \psi (k, \Delta t) = 1.
\label{F2+} 
\end{equation} 
We consider $\hat \psi (k, \Delta t)$ as the average time-scale- and wave-vector-dependent transition ``probability" or the characteristic function of the stochastic process in Eq.~(\ref{2}). In general, $\hat \psi (k, \Delta t)$ can be due to many independent, co-existing processes, each characterized by its own, ``partial" transition probability, $\psi_j (k, \Delta t)$, $j=1,\dots n$, making it possible to expand
\begin{equation}
\hat \psi (k, \Delta t) = \prod_{j=1}^n \hat \psi_j (k, \Delta t).
\label{Prod} 
\end{equation} 
We should stress that, by their definition as Fourier integrals, $\hat \psi_j (k, \Delta t)$ are given by complex functions of the wave vector $k$, and that their appreciation as ``probabilities" has the only purpose of factorizing in Eq.~(\ref{Prod}). Even so, with the aid of Eq.~(\ref{F2}) above, this factorized form is justified via the asymptotic matching procedure in the limit $k\rightarrow 0$. In practice, aiming at the prospective fluid and plasma applications, it is sufficient to address a simplified version of Eq.~(\ref{Prod}), where two co-existing key processes are included $-$ one corresponding to a white noise-like process, which we shall mark by the index $L$; and the other one, corresponding to a regular convection process, such as a zonal flow or similar, which we shall mark by the index $R$. We have, accordingly,  
\begin{equation}
\hat \psi (k, \Delta t) = \hat \psi_L (k, \Delta t) \hat \psi_R (k, \Delta t).
\label{Prod2} 
\end{equation}
These settings correspond to a set of Langevin equations
\begin{equation}
dx/dt = v;~dv/dt = -\nu v + F_R + F_L (t),
\label{Lvin} 
\end{equation}
where $\nu$ is the fluid viscosity; $F_R$ is the regular force; and $F_L (t)$ is the fluctuating (noise-like) force. We take $F_L (t)$ to be a white L\'evy noise with L\'evy index $\mu$ ($1 < \mu\leq 2$). By white L\'evy noise $F_L (t)$ we mean a stationary random process, such that the corresponding motion process, i.e., the time integral of the noise, $L (\Delta t) = \int_t^{t+\Delta t} F_L (\tau) d\tau$, is a symmetric $\mu$-stable process with stationary independent increments and the characteristic function  
\begin{equation}
\hat \psi_L (k, \Delta t) = \exp (-D_\mu |k|^\mu \Delta t) \sim 1 - D_\mu |k|^\mu \Delta t.
\label{GCLT} 
\end{equation}
The last term gives an asymptotic inverse-power distribution of jump lengths $\lambda (x) \sim |x|^{-1-\mu}$. The constant $D_\mu$ constitutes the intensity of the noise. Note that the general condition in Eq.~(\ref{F2+}) is clearly satisfied. As is well-known, the characteristic function in Eq.~(\ref{GCLT}) generates L\'evy flights \cite{Klafter,Report}. 

Focusing on the regular component of the force-field, $F_R$, it is convenient to choose the corresponding transition probability in the form of a plane wave, i.e.,
\begin{equation}
\hat \psi_R (k, \Delta t) = \exp (iuk\Delta t) \sim 1 + iuk\Delta t.
\label{Plane} 
\end{equation}
Here, $u$ is the speed of the wave, which is decided by convection. One evaluates this speed by neglecting the term $dv/dt$ in Langevin Eqs.~(\ref{Lvin}) to yield $u=F_R/\nu$. Putting all the various pieces together, one readily obtains 
\begin{equation}
\hat \psi (k, \Delta t) = \exp (-D_\mu |k|^\mu \Delta t + ik F_R \Delta t / \nu).
\label{Tog} 
\end{equation} 
The next step is to substitute this into Eq.~(\ref{3}), and to allow $\Delta t \rightarrow 0$. Then, Taylor expanding on the left- and right-hand sides in powers of $\Delta t$, and keeping first non-vanishing orders, in the long-wavelength limit $k\rightarrow 0$ it is found that 
\begin{equation}
\frac{\partial}{\partial t} \hat f (k, t) = \left[-D_\mu |k|^\mu + ik F_R / \nu \right] \hat f (k, t).
\label{9} 
\end{equation} 
When inverted to configuration space, the latter equation becomes
\begin{equation}
\frac{\partial}{\partial t} f (x, t) =  \left[D_\mu \frac{\partial^\mu}{\partial |x|^\mu} - \frac{1}{\nu} \frac{\partial}{\partial x} F_R \right] f (x, t),
\label{Inv} 
\end{equation} 
where the symbol $\partial^\mu / \partial |x|^\mu$ is defined by its Fourier transform as  
\begin{equation}
\hat \mathrm{F} \Big\{\frac{\partial^\mu}{\partial |x|^\mu}f (x, t)\Big\} = -|k|^\mu \hat f (k,t),
\label{Def} 
\end{equation} 
and we have followed the usual convention \cite{Klafter} of suppressing the imaginary unit in Fourier space. In the foundations of fractional calculus \cite{Samko} it is shown that, for $1 <\mu < 2$, 
\begin{equation}
\frac{\partial^\mu}{\partial |x|^\mu} f (x, t) = \frac{1}{\Gamma_\mu}\frac{\partial^2}{\partial x^2} \int_{-\infty}^{+\infty}\frac{f (x^\prime, t)}{|x-x^\prime|^{\mu - 1}} dx^\prime,
\label{Def+} 
\end{equation} 
where $\Gamma_\mu = - 2\cos(\pi\mu/2)\Gamma(2-\mu)$ is a numerical normalization parameter. One sees that $\partial^\mu / \partial |x|^\mu$ is an integro-differential operator, which has the analytical structure of ordinary space differentiation acting on a Fourier convolution of the function $f (x,t)$ with a power-law. It interpolates between a pure derivative and a pure integral, and is often referred to as the fractional Riesz/Weyl operator. By its definition, the Riesz/Weyl operator can conveniently be considered as a normalized sum of left and right Riemann-Liouville derivatives on the infinite axis. It is this operator, which incorporates the nonlocal properties of the transport. In the Gaussian limit $\mu = 2$, the Riesz/Weyl operator reduces to the conventional Laplacian, so that local behavior is recovered. Relating $F_R$ to an external potential field, $F_R = -V^{\prime}(x)$, we are led to the following fractional Fokker-Planck equation, or FFPE (e.g., Refs. \cite{Gonchar,Jesper,Chechkin,Fog}; reviewed in Ref. \cite{Klafter}) 
\begin{equation}
\frac{\partial}{\partial t} f (x, t) =  \left[D_\mu \frac{\partial^\mu}{\partial |x|^\mu} + \frac{1}{\nu} \frac{\partial}{\partial x} V^{\prime}(x)\right] f (x, t).
\label{FFPE} 
\end{equation} 
In writing FFPE with the spatial dependence in $V^{\prime}(x)$ we have also assumed that the scales of the $F_R$ variation are smooth compared with the fluctuation noise-like scales. 

Equation~(\ref{FFPE}) can be extended, so that it includes local transport due to e.g., collisions, in addition to the nonlocal transport processes discussed above. The key step is to observe that collisions will generate a white noise of the Brownian type whose characteristic function is just a Gaussian, and is obtained from the general L\'evy form, Eq.~(\ref{GCLT}), for $\mu\rightarrow 2$. We note in passing that the Gaussian law, too, belongs to the class of stable distributions, but it will be the only one to produce finite moments at all orders. When the L\'evy and Brownian noises are included as independent elements to the dynamics, the transition probability in Eq.~(\ref{Prod}) will again factorize, and will acquire, in addition, a Gaussian factor $\hat \psi_G (k, \Delta t) = \exp (-D k^2 \Delta t)$, where $D$ has the sense of collisional diffusion coefficient. Then Eq.~(\ref{Tog}) will generalize to   
\begin{equation}
\hat \psi (k, \Delta t) = \exp (-D_\mu |k|^\mu \Delta t -D k^2 \Delta t + ik F_R \Delta t / \nu),
\label{CLT+} 
\end{equation}  
from which a FFPE involving both the Riesz/Weyl and the usual Laplacian operators 
\begin{equation}
\frac{\partial}{\partial t} f (x, t) =  \left[D_\mu \frac{\partial^\mu}{\partial |x|^\mu} + D \frac{\partial^2}{\partial x^2} + \frac{1}{\nu} \frac{\partial}{\partial x} V^{\prime}(x) \right] f (x, t)
\label{FFPE+} 
\end{equation} 
can be deduced for small $k$. In the applications, it is convenient to think of the L\'evy noise as occurring above a certain threshold, so that $D_\mu$ contains a step-like dependence switching between local (e.g., collisional, as well as Gaussian quasi-linear) and nonlocal transport.  

\section{To be, or not to be \cite{Shak}} We have seen in the above that the statistical case of nonlocal transport stems from a driving noise-process of the L\'evy type, whereas regular convection acts as to introduce an external potential field to L\'evy flights. On the one hand, this spotlights the basic physics implications of the fractional derivative operator occurring in FFPE, Eq.~(\ref{FFPE}). On the other hand, it casts doubts on its relevance to the classical picture of drift-wave- and two-dimensional fluid turbulence, as well as to the classical picture of SOC. The main elements of concern consist in the following. In SOC, one is interested in how long-time correlated dynamics will develop via local couplings between the many degrees of freedom leading to complex patterns \cite{Bak,Ash2013}. Then the assumed next-neighbor character of lattice interactions in the vicinity of criticality will be incompatible with nonlocal space differentiation, so that the correlations that are long-ranged enter through nonlocal differentiation over the time, rather than the space, variable, thus preserving the local structure of the Laplacian (Ref. \cite{NJP}; Chapter 4 of Ref. \cite{Ash2013}). Further concerns come from non-observation, over a statistically significant range, of algebraic tails in direct numerical simulations of electrostatic drift-wave turbulence, also supported by the available experimental evidence. A particle going with a strongly turbulent flow will experience a sequence of flights, and we can reasonably expect that the distribution of flight lengths will approximately follow the average particle flux distribution. The latter is obtained as a flux-surface integral of the convected density, i.e., $\Gamma = \int d\sigma \left[\tilde n u_{E\times B}\right]/\int d\sigma$, where $\tilde n$ and $u_{E\times B}$ are the density fluctuations and the $E\times B$ velocity, respectively. Then a statistics of the L\'evy type will imply that the probability density function of the averaged transport $\Gamma$ will exhibit an algebraic tail, i.e., $\lambda (\Gamma) \sim \Gamma^{-1-\mu}$, whereas the simulations have revealed an exponential tail \cite{Basu,PLA,Garcia,Garcia+}.     

Based on this reasoning, we are led to infer that a nonlocal FFPE is neither consistent with the classical drift-wave- or two-dimensional fluid turbulence approach nor a SOC approach built on the assumptions of locality and next-neighbor interactions. Thus, a more intricate gateway for nonlocal transport should be agreed. Here, we suggest that nonlocality comes into play as a result of amplification, involving SOC avalanches in the presence of inverse energy cascade due to turbulence. This idea leads directly to the L\'evy statistics, as we now proceed to show.    

\section{Turbulent amplification process} In the combined SOC-turbulence scenario, which we consider, a guiding role is attributed to the usual picture of eddies and eddy-induced transport associated with plasma instabilities. A good candidate to understand and explain the phenomenon from first principles is at present drift-wave turbulence modeled by the well-known Hasegawa-Wakatani (HW) equations, as detailed numerical investigations show \cite{Basu,PLA,JJR}. Nonlinearly, in a magnetic confinement geometry, the eddy-induced transport reduces the slope of the average profile where the vorticity is maximal, and, at the same time, steepens it in its nearby vicinity, thus increasing the instability in the next radial location. This displacement of the instability is an avalanche in that the step in the gradient moves radially outward, creating an unstable propagating front.\footnote{A report on observation and quantitative characterization of avalanche events in a magnetically confined plasma can be found in Ref. \cite{Politzer2}. It was argued that the evidence that avalanche events are present in the plasma is ``strong," and that the observations are qualitatively similar to results of modeling calculations based on drift-wave turbulence.} However, because drift-wave turbulence is essentially two-dimensional, there exists an inverse cascade of the energy which is associated with the phenomena of eddy merging and the formation of large-scale coherent structures in the strongly turbulent flow. To this end, the propagation of the unstable front becomes a combined effect due to the next-radial generation of the off-spring eddies and their merging with the ever-growing mother eddy. One sees that turbulence will act as to amplify the avalanches by fueling them with more free energy via the inverse cascade. The process will stop when excess energy and particles are eventually let out through boundaries, thus reducing the slope of the average profile on the system-size scales. 

The amplification occurs, when the eddy turnover time, $\tau_{\rm turn}$, is small compared with the instability growth time. For the purpose of formal ordering, we require $\tau_{\rm turn}\ll \gamma_{\rm L}^{-1}$, where $\gamma_{\rm L}$ is the linear growth rate. In drift-wave turbulence, the so-called Rhines length, $\lambda_{\rm Rh}$, determines the upper bound on the size of vortical structures. The Rhines length, originally introduced in geophysical fluid turbulence \cite{Rhines}, and later applied to drift-wave turbulence \cite{Nau}, designates the spatial scale separating vortex motion from drift wave-like motion. For the HW system it is obtained as $\lambda_{\rm Rh} \propto \sqrt{u_{E\times B}}$, leading to a characteristic turnover time $\tau_{\rm turn} \sim \lambda_{\rm Rh} / u_{E\times B} \propto 1/ \sqrt{u_{E\times B}}$. Next, the instability growth rate $\gamma_{\rm L}$ is proportional with nonadiabaticity of the fluctuations. The nonadiabaticity parameter, $\delta$, characterizes deviation between the potential and the density fluctuations in the HW model. It constitutes an internal drive for the turbulence, offering a maximum growth rate $\gamma_{\rm L} \approx \delta / 8$ \cite{JJR}. It is worth noting that $\delta$ absorbs the parameters of parallel dynamics; it contains via the electron-ion collisional frequency the parallel resistivity. One sees that the conditions for turbulent amplification of the avalanches are favored in the regime of nearly adiabatic fluctuations on the one hand, and by stronger nonlinearities taking place, on the other hand. These conditions of slow driving combined with the guiding role of boundary dissipation constitute a typical set-up for dynamical systems exhibiting SOC \cite{Ash2013}. At this point, the propagation of unstable fronts due to the processes of eddy merging and interactions acquires the typical signatures of the avalanching dynamics of the SOC type. The existence of outer boundary is very important in this scenario, as it introduces a feedback of the particle and energy loss processes on the dynamical state of the turbulent system. The observational consequence of feedback lies in the fact that the average gradient controlling the instabilities is attracted to the critical value, where the system is essentially very sensitive to perturbations. Nonlinearly, when a saturated turbulence state is approached, the initial fine-scale signatures of the fluctuations are washed out by amplification. When this occurs, the dynamics are dominated by the bulk-average nonlinearities, and not anymore by multi-scale features of fluctuations in the HW regime, enabling appreciable departures from the state of marginal stability. Analysis of this behavior requires global models accounting for a self-consistent evolution of the background density profiles on an equal footing with the multi-scale fluctuation dynamics.  

We suggest that the mechanism of turbulent amplification of the SOC avalanches is responsible for the occurrence of large-amplitude transport events in plasma confinement, associable with large intermittent bursts of transport \cite{Ippolito}. We expect these bursts to generate algebraically decaying tails in the distributions of energy and particle flux, thus giving rise to statistics of the L\'evy type.    

This suggestion finds its validation in the general properties of log-normal behavior. It is noticed that the flux-surface averaged transport, $\Gamma = \int d\sigma \left[\tilde n u_{E\times B}\right]/ \int d\sigma$, is to lowest order positive definite \cite{PLA}, and that it shows a multiplicative character in that it involves a product of two random quantities, the fluctuating particle density, $\tilde n$, and the fluctuating $E\times B$ velocity. Then the central limit theorem will imply that it is the logarithm of the averaged flux, which is normally distributed. Indeed it is found in direct numerical simulations of the HW model that the probability distribution functions of the flux-surface averaged transport are generally in good agreement with a log-normal distribution (see Fig.~1). These results also generalize to electromagnetic drift-Alfv\`en turbulence with magnetic field curvature effects, as well as to magnetohydrodynamic edge plasma turbulence \cite{PLA,Garcia,Garcia+}. As Montroll and Shlesinger realized \cite{Montroll}, an initially log-normal distribution will change to a distribution without well-defined moments when/if its mean value is unlimitedly amplified by some process. With the aid of a recursion relation it was argued that the new distribution that allows for these amplifications has a nonanalytic part leading to the Pareto-L\'evy tail in the sense of a $\mu$-stable L\'evy motion and generalized central limit theorem. Interestingly, the above authors motivated their study with explanation of $1/f$ noise and with examples involving hierarchical random processes with subordination. They suggested that, when subordination occurs in many orders, the distribution function of successes in the primary order is log-normal. Using here that the inverse energy cascade, characteristic of two-dimensional fluid turbulence, acts as to introduce a ``subordination" into the hierarchy of turbulent eddies, parametrized by their wave number, and that the avalanches themselves emerge from same turbulence vortex motions provided just that there is a time scale separation, $\tau_{\rm turn}\ll \gamma_{\rm L}^{-1}$, one readily concludes that the amplification processes taking place will naturally generate the wanted Pareto-L\'evy inverse-power-law tail, $\lambda (\Gamma) \sim \Gamma^{-1-\mu}$, consistently with the flight length distribution dictated by Eq.~(\ref{GCLT}), and the $\mu$ exponent defining the hierarchy of the amplification. It is this L\'evy motion, obtained through the amplification of log-normal behavior, which we charge to generate, via the corresponding white L\'evy noise $F_L (t)$, a nonlocal FFPE. 

The net result of the discussion above is that space-fractional derivatives enter FFPE very nontrivially in that neither a classical SOC nor classical turbulence models can readily accommodate them as such. We have seen that an amplification process is needed to validate nonlocal models of anomalous transport. The physics picture of the amplification has referred to a synergetic coupling between SOC and drift-wave turbulence phenomena in the presence of nonlinear feedbacks at the plasma edge. 
Because of amplification, we expect a steeper drop-off in the energy spectrum as compared to respective fluid- and drift-wave- counterparts. Taking this idea to its extreme limit, we obtain for small wave vectors $E(k)\sim k^{-3}$. In this connection, we should stress that the avalanching transport is triggered by some radial dependence in the profiles, and, when account is taken for the inverse cascade, by boundary feedbacks, so that the assumptions of constant energy transfer and of infiniteness of the system, resulting in the fluid-like $-5/3$ behavior, do not really apply here. Whereas the boundaries can, in the thermodynamic limit, be assumed at infinity, they are an essential key element to the model as they introduce a feedback dynamics generating SOC.  
\begin{figure}
\includegraphics[width=0.49\textwidth]{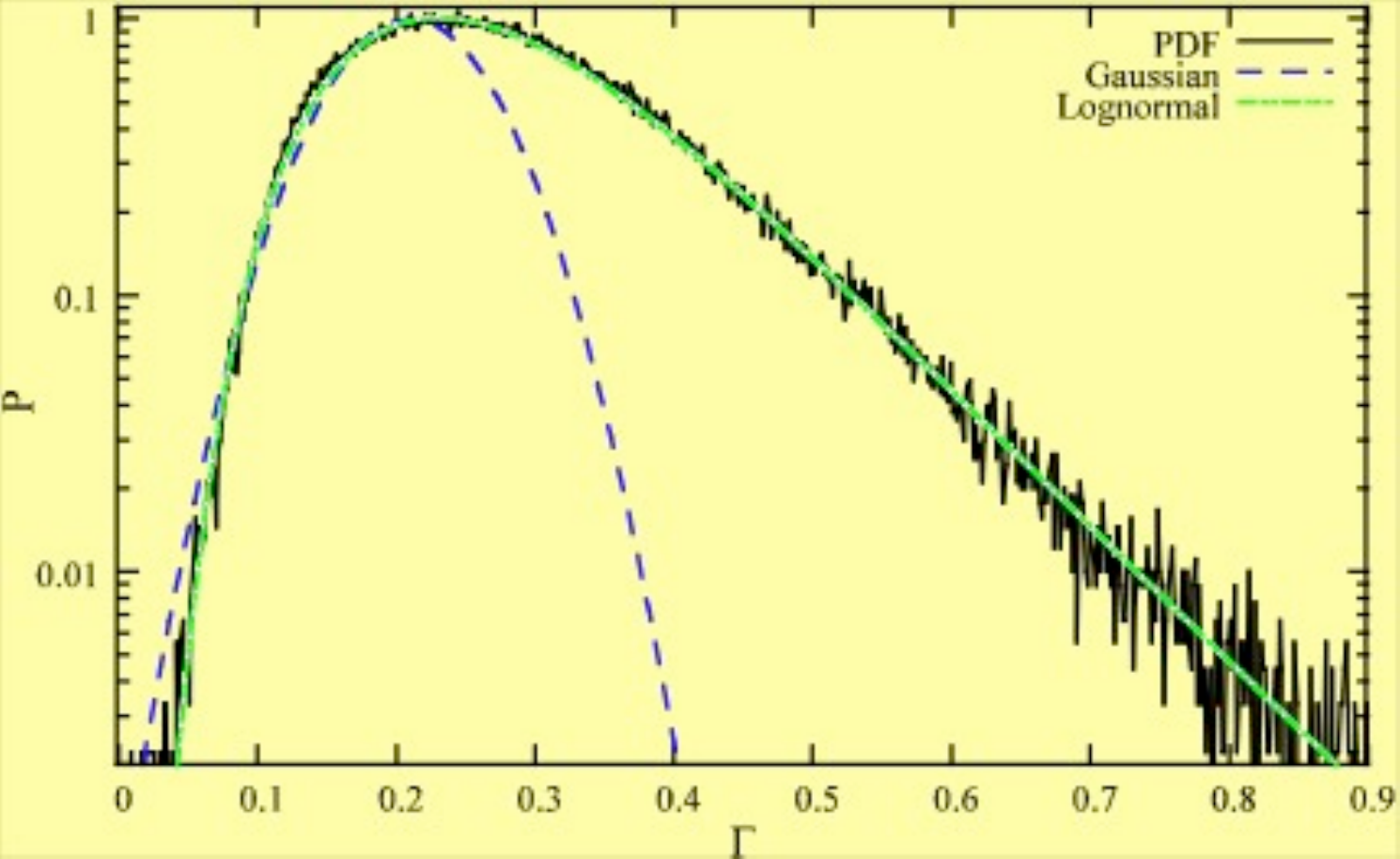}
\caption{\label{} Probability distribution functions of the flux-surface averaged plasma flux compared with a log-normal and Gaussian distributions from a numerical solution of the HW equation. Adapted from Ref. \cite{PLA}.}
\end{figure}
The operation of a negative feedback mechanism, other than providing a route to SOC phenomena, guarantees the necessary stiffness to profiles near a marginally stable state. It will also explain some asymmetry \cite{Pulse,Pulse+} between the propagation of perturbations due to heat modulation and cold pulses. For $x > x_s$, where $x_s$ denotes the location of ion cyclotron resonance heating power deposition, heat waves and pulses propagate fast. However, for $x < x_s$, the heat wave slows down and is damped, but the cold pulses still travel fast. The explanation lies in the fact that the application of a cold pulse to the plasma edge steepens the gradient of the average profile, thus turning it into the unstable (supercritical) domain. Then the system responds by plasma instabilities and the phenomena of turbulent amplification of the fluxes. So, the transport is nonlocal, and the transport problem for the cold pulse is essentially a L\'evy flight problem. By contrast, the application of heat power modulation introduces some sort of knee to the profile. Indeed the profile becomes steeper (and thus unstable) in the range $x > x_s$ where behavior is supercritical involving nonlocality, and flatter in the range $x < x_s$ which is subcritical and which damps the wave. 

\section{Conclusions} We propose a combined model of nonlocal transport which brings the notion of inverse energy cascade, typical of drift-wave- and two-dimensional fluid turbulence, in contact with the ideas of avalanching behavior, characteristic of SOC. In this spirit, we suggest that SOC and turbulence mechanisms couple in magnetically confined plasma, and operate on essentially an equal footing to introduce nonlocal features into dynamics. This occurs via the phenomena of turbulent amplification of avalanches of the SOC type, giving rise to large intermittent bursts of transport. The SOC properties step in via a nonlinear feedback of edge behavior on the dynamical turbulent state of the core plasma. We have seen that SOC is not really an alternative to turbulence in this description. It operates as a dynamically induced nonlinear twist in a basically turbulent medium. This nonlinearity is implicitly present in the value of the fractional exponent $\mu$ used to generalize the Laplacian. It leads to a nontrivial situation, in which the transport equation, the FFPE, is formally linear, with feedback nonlinearities absorbed by the fractional order of space differintegration. 

Based on this reasoning, we expect that, when interchange turbulence for the scrape-off layer region is included\cite{PLA,Garcia,Garcia+}, the phenomena of amplification will increase in importance. The nonlocal model implies that the interaction between SOC- and turbulence-associated processes will manifest itself as algebraically decaying tails on top of typically log-normal-like probability distributions of the average heat and particle flux. The idea of SOC-turbulence coupling supports the general way of thinking that plasmas of fusion interest operate as a complex self-organized system with nonlinear feedbacks between the many co-existing dynamical processes involved \cite{ChZo}. The comprehension of complexity and self-organization of a plasma in magnetic confinement geometry poses a theoretical challenge. Here, we have further advanced these topics by discussing an inherent interconnection between the dynamical properties due to turbulence, self-organized criticality, and boundary feedbacks. In this spirit, we have provided some mathematical tools for consistency analysis of the transport scenarios behind these approaches. A hard result here is the derivation of space-fractional FFPE for a random walker driven by a white L\'evy noise. It was argued that L\'evy motions were generated universally through an amplification process \cite{Montroll}, and that their characteristics were instrumental in describing nonlocality. We believe that the significance of this process goes beyond the plasma applications. In the specific physics context of SOC-turbulence coupling, the exponent of nonlocal differentiation, $\mu$, could be thought of as defining the typical anomalous scaling of lifetime of the avalanches with their size, i.e., $t\sim |x|^\mu$, characteristic \cite{Castillo2} of the low confinement mode plasma. Essentially the same scaling relation is obtained from major SOC models where it derives from the assumptions of locality and next-neighbor interactions. In those settings the $\mu$ exponent, which is model dependent, determines the universality class of SOC \cite{Maslov}. There is no indication that our mixed SOC-turbulence model falls within the known universality classes. In this respect, the observation of a power-law, $\lambda (\Gamma) \sim \Gamma^{-1-\mu}$, does not really imply SOC. It does imply a SOC-turbulence coupling instead. It's not the taxonomy we are pointing at: The basic phenomenon is generic.

\acknowledgments
We thank V. Naulin for discussion and comments to the manuscript. 
This work was supported by the Euratom Communities under the contract of Association between Euratom/ENEA. Partial support was received from the ISSI project ``SOC and Turbulence" (Bern, Switzerland).

\end{document}